\documentstyle[12pt]{article}

\newcommand{\beq}{\begin{equation}}
\newcommand{\beqn}{\begin{displaymath}}
\newcommand{\zen}{\end{equation}}
\newcommand{\zenn}{\end{displaymath}}
\def\pnot{\mbox{${\not{\hbox{\kern-3.0pt$p$}}}$}}
\def\qnot{\mbox{${\not{\hbox{\kern-2.0pt$q$}}}$}}
\def\enot{\mbox{${\not{\hbox{\kern-2.0pt$e$}}}$}}
\def\knot{\mbox{${\not{\hbox{\kern-2.0pt$k$}}}$}}
\def\rnot{\mbox{${\not{\hbox{\kern-2.0pt$r$}}}$}}
\def\r'not{\mbox{${\not{\hbox{\kern-2.0pt$r'$}}}$}}

\def\fun#1#2{\lower3.6pt\vbox{\baselineskip0pt\lineskip.9pt\ialign
{$\mathsurround=0pt#1\hfil##\hfil$\crcr#2\crcr\sim\crcr}}}

\newcommand{\lsim}{\ \raise -2.truept\hbox{\rlap{\hbox{$\sim$}}
\raise5.truept \hbox{$<$}\ }}
\newcommand{\gsim}{\ \raise -2.truept\hbox{\rlap{\hbox{$\sim$}}
\raise5.truept\hbox{$>$}\ }}

\begin{document}

%\newpage
%\setcounter{page}{1}
\begin{titlepage}
\hskip 12cm \vbox{\hbox{BUDKERINP/96-35}\hbox{CS-TH 3/96}\hbox{May 1996}}
\vskip 0.3cm
\centerline{\bf GLUON REGGE TRAJECTORY IN THE TWO-LOOP 
APPROXIMATION$^{~\ast}$}
\vskip 0.8cm
\centerline{  V.S. Fadin$^{\dagger}$}
\vskip .1cm
\centerline{\sl Budker Institute for Nuclear Physics}
\centerline{\sl and Novosibirsk State University, 630090 Novosibirsk,
Russia}
\vskip .4cm
\centerline{  R. Fiore$^{\ddagger}$}
\vskip .1cm
\centerline{\sl  Dipartimento di Fisica, Universit\`a della Calabria}
\centerline{\sl Istituto Nazionale di Fisica Nucleare, Gruppo collegato di
Cosenza}
\centerline{\sl Arcavacata di Rende, I-87030 Cosenza, Italy}
\vskip .4cm
\centerline{  M.I. Kotsky$^{\dagger}$}
\vskip .1cm
\centerline{\sl Budker Institute for Nuclear Physics}
\centerline{\sl  630090 Novosibirsk, Russia}
\vskip 0.8cm
\begin{abstract}
Evaluating two-loop integrals in the transverse momentum space we obtain 
the trajectory of the Reggeized gluon in QCD in an explicit form in the 
two-loop approximation. It is presented as an expansion in powers of 
$(D-4)$ for the space-time dimension $D$ tending to the physical value 
$D=4$. As the result of a remarkable cancellation the third order pole in 
$D-4$ disappears in the trajectory and the expansion starts with $(D-4)^{-2}$.
\end{abstract}
\vskip .5cm
\hrule
\vskip.3cm
\noindent

\noindent
$^{\ast}${\it Work supported in part by the Ministero italiano 
dell'Universit\`a e della Ricerca Scientifica e Tecnologica, in part by the 
EEC Programme ``Human Capital and Mobility", Network ``Physics at High Energy 
Colliders", contract CHRX-CT93-0357 (DG 12 COMA), in part by INTAS grant 
93-1867 and in part by the Russian Fund of Basic Researches.}
\vfill
%$^{a}${{\it Permanent address}: Budker Institute for Nuclear Physics and 
%Novosibirsk State University, Novosibirsk, Russia} 
\vskip .2cm
$ \begin{array}{ll}
^{\dagger}\mbox{{\it email address:}} &
 \mbox{FADIN, KOTSKY~@INP.NSK.SU}\\
\end{array}
$

$ \begin{array}{ll}
^{\ddagger}\mbox{{\it email address:}} &
  \mbox{FIORE~@FIS.UNICAL.IT}
\end{array}
$
\vfill
\end{titlepage}
\eject
\textheight 220mm
\baselineskip=24pt
{\bf 1. INTRODUCTION}

Dynamics of semihard processes (i.e. processes with the c.m.s. energy 
$\sqrt{s}$ much larger than a typical virtuality Q) is one of the most 
interesting problems of modern high energy physics. There are a lot of 
papers devoted to the theoretical description of the processes (see, for 
example, Refs. [1,2] and references therein). Nevertheless, the 
present situation here is not quite satisfactory, even in the framework 
of perturbative QCD which, presumably, can be applied to such processes 
\cite{GLR}. The BFKL equation \cite{FKL} used for the calculation of the 
small $x=\frac{Q^2}{s}$ behaviour of cross-sections is responsible for the 
leading $\ln(1/x)$ terms only. To define a region of its applicability as 
well as to fix a scale of virtualities of the running coupling constant 
$\alpha_s(Q^2)$ in the equation one has to know radiative corrections to 
the kernel of the equation. The crucial point in the programme of 
calculation of the next-to-leading corrections \cite{LF} is the gluon 
Reggeization in QCD. The Reggeization formed a basis of the original 
derivation of the BFKL equation \cite{LF} and was proved in the leading 
$\ln(1/x)$ approximation for elastic and inelastic processes \cite{BLF}. 
In this approximation the gluon trajectory 
\beqn
j(t) = 1+\omega(t)~,
\zenn
\beq
\omega(t) = \omega^{(1)}(t)+\omega^{(2)}(t)+\cdots
\label{z1}
\zen
can be written as an integral in the transverse momentum space:
\beq
\omega^{(1)}(t) = \frac{g^2t}{(2{\pi})^{(D-1)}}\frac{N}{2}
\int \frac{d^{D-2}k}{{\vec k^2}{(\vec q- \vec k)^2}}~,~~~~~~
t = -\vec q^{~2}~,
\label{z2}
\zen
corresponding to a simple Feynman diagram of $(D-2)$-dimensional field 
theory. Here and below $N$ is the number of colours ($N=3$ for QCD). The 
integral is divergent in the physical case $D=4$. In fact one gets
\beq
\omega^{(1)}(t) = -\bar g^2(\vec q^{~2})^{\mbox{\normalsize $\epsilon$}} 
~\frac{2}{\epsilon}
%\frac{\Gamma^2 (1+\epsilon )}{\Gamma (1+2\epsilon )}~.
\frac{\Gamma^2 (1+\epsilon )}{\Gamma (1+2\epsilon )}~,
\label{z3}
\zen
%Here and below we put 
where
\beq
\epsilon = \frac{D}{2}-2~,~~~~~~~~\bar g^2 = \frac{g^2N}{(4{\pi})^
{\frac{D}{2}}}\Gamma(1-\epsilon)~.
\label{z4}
\zen
This divergency is effectively cancelled in the kernel of the BFKL equation 
\beq
{\cal K}(\vec q,\vec q^{~'}) = 2\omega(t)\delta(\vec q-\vec q^{~'})+
{\cal K}_{real}(\vec q,\vec q^{~'})~,
\label{z5}
\zen
\beq
{\cal K}^{(1)}_{real}(\vec q,\vec q^{~'}) = \frac{\bar g^2}{\pi^{1+
\mbox{\normalsize $\epsilon$}}
\Gamma(1+\epsilon)}\frac{4}{(\vec q- \vec q^{~'})^2}~,
\label{z6}
\zen
which is not singular at $D=4$ for integration with any smooth function of
$\vec q^{~'}$.
\vskip.03cm
In the next-to-leading approximation Reggeization was checked and the 
correction to the trajectory was presented in the form of two-loop 
integrals in the transverse momentum space [6,7]. However, it appears that 
we have to calculate the integrals because of increased infrared 
divergency of separate terms in the correction to the kernel which 
diverge as $(D-4)^{-3}$. So, contrary to what 
happened in the leading order, where the divergency cancellation was organized 
by a suitable rearrangement of virtual and real contributions to the kernel 
without going to $D \neq 4$, in the correction one hardly can hope to do it. 
Therefore, we need to calculate the correction to the trajectory in an 
explicit form. In the present paper we solve this problem.

\vskip 0.3cm

{\bf 2. REPRESENTATION OF THE QUARK AND GLUON \\
\hspace*{1.0cm} CORRECTIONS}

The two-loop correction contains quark as well as gluon contributions.
\vskip.03cm
Let us first consider the quark contribution. Of course, here the case of 
light quarks is the most important one. Due to the cancellation of infrared 
terms in the equation masses of the light quarks should not affect the BFKL 
dynamics, so we can put them equal to zero. Therefore, according to Refs. 
[6,7] this contribution can be written as
\beq
\omega_q^{(2)}(t) = \frac{\bar g^4 \vec q^{~2}}{\pi^{1+
\mbox{\normalsize $\epsilon$}}\Gamma(1-
\epsilon)}\frac{4n_f}{N\epsilon}\int \frac{d^{D-2}q_1}{\vec q_1^{~2}
(\vec q- \vec q_1)^2}\int_0^1dx(x(1-x))^{1+
\mbox{\normalsize $\epsilon$}}\left((\vec q^{~2})^
{\mbox{\normalsize $\epsilon$}}-2(\vec q_1^{~2})^
{\mbox{\normalsize $\epsilon$}}\right)~.
\label{z7}
\zen
where $n_f$ is the number of light flavours. Making use of the generalized
Feynman parametrization
\beq
\prod_{i=1}^n a_i^{-\alpha_i} = \frac{\displaystyle{\Gamma\left(\sum_{i=
1}^{n}\alpha_i\right)}}{\displaystyle{\prod_{i=1}^{n}\Gamma(\alpha_i)}}
\left(\prod_{i=1}^n\int_0^1dx_ix_i^{\alpha_i-1}\right)\frac{
\displaystyle{\delta(1-\sum_{i=1}^{n}x_i)}}{\displaystyle{\left(
\sum_{i=1}^{n} a_ix_i\right)^{\displaystyle{\sum_{i=1}^{n}\alpha_i}}}}
\label{z8}
\zen
the integration in Eq. (\ref{z7}) can be performed without any difficulty 
and leads to
\beq
\omega_q^{(2)}(t) = \bar g^4(\vec q^{~2})^{2
\mbox{\normalsize $\epsilon$}}\frac{4n_f}{N\epsilon}
\frac{\Gamma^2(2+\epsilon)}{\Gamma(4+2\epsilon)}
\left[\frac{2}{\epsilon}\frac{\Gamma^2(1+\epsilon)}{\Gamma(1+2\epsilon)}-
\frac{3}{\epsilon}\frac{\Gamma(1-2\epsilon)}{\Gamma^2(1-\epsilon)}
\frac{\Gamma(1+\epsilon)\Gamma(1+2\epsilon)}{\Gamma(1+3\epsilon)}\right]~.
\label{z9}
\zen
For the case $\epsilon \rightarrow 0$ Eq. (\ref{z9}) gives
\beq
\omega_q^{(2)}(t) = -\bar g^4(\vec q^{~2})^{2\mbox{\normalsize $\epsilon$}}
\frac{2n_f}{3N}\frac{1}{\epsilon^2}\left[1-\frac{5}{3}\epsilon-
\left(\frac{\pi^2}{3}-\frac{28}{9}\right)\epsilon^2\right]~.
\label{z10}
\zen
\vskip.03cm
The gluon contribution is more complicated. It reads \cite{FFK}
\beqn
\omega^{(2)}_g(t) = (\vec q^{~2})^{2
\mbox{\normalsize $\epsilon$}}\left\{\left[\psi(1)+2
\psi(\epsilon)-\psi(1-\epsilon)-2\psi(1+2\epsilon)\right.\right.
\zenn
\beq
\left.\left. +\frac{1}{(1+2\epsilon)}\left(\frac{1}{\epsilon}+\frac{1+
\epsilon}{2(3+2\epsilon)}\right)\right](I_1-2J_1)-2J_2+2J_3-I_2\right\}~,
\label{z11}
\zen
where $\psi(x)$ is the logarithmic derivative of the gamma function, 
$\psi(z)$=$d\ln\Gamma(z)$/$dz$, and the quantities $I_i$, $J_i$ are 
two-loop integrals in the transverse momentum space. They can be written 
in the following form:
\beq
I_i = {\left(\frac{g^2N}{2(2\pi)^{D-1}}\right)}^2\int\frac{d^{(D-2)}q_1
d^{(D-2)}q_2(\vec q^{~2})^{6-D}}{\vec q_1^{~2}(\vec q_1-\vec q)^2
\vec q_2^{~2}(\vec q_2-\vec q)^2}a_i~,
\label{z12}
\zen
with
\beq
a_1 = 1~,~~~~~~~a_2 = \ln\left(\frac{(\vec q_1-\vec q_2)^2}
{\vec q^{~2}}\right)
\label{z13}
\zen
and
\beq
J_i = {\left(\frac{g^2N}{2(2\pi)^{D-1}}\right)}^2\int\frac{d^{(D-2)}q_1
d^{(D-2)}q_2(\vec q^{~2})^{5-D}}{\vec q_1^{~2}\vec q_2^{~2}(\vec q-\vec q_1
-\vec q_2)^2}b_i~,
\label{z14}
\zen
with
\beq
b_1 = 1~,~~~~~~b_2 = \ln\left(\frac{(\vec q-\vec q_2)^2}{\vec q^{~2}}\right)~,
~~~~~~b_3 = \ln\left(\frac{\vec q_2^{~2}}{\vec q^{~2}}\right)~.
\label{z15}
\zen
Let us note that because of the logarithmic factors (coming from the 
integration over longitudinal momentum components) the integrals 
$I_i$ and $J_i$, with $i\geq2$, do not correspond to usual Feynman diagrams 
of $D-2$-dimensional field theory.
\vskip.03cm
The integral $I_1$ is given by the square of the integral for the leading 
contribution to the trajectory (\ref{z2}) and reads
\beq
I_1 = \bar g^4\left(\frac{2}{\epsilon}\frac{\Gamma^2(1+\epsilon)}
{\Gamma(1+2\epsilon)}\right)^2~.
\label{z16}
\zen
The calculation of the integral $J_1$ is not much more complicated than 
that of $I_1$. It can be performed by the subsequent integration over 
$\vec q_1$ (using the Feynman parametrization to join the denominators 
$\vec q_1^{~2}$ and $(\vec q-\vec q_1-\vec q_2)^2)$ and $\vec q_2$. At the 
second step a non-integer power of $(\vec q_2-\vec q)^2$ appears in the 
denominator, so that we need to use the generalized parametrization 
(\ref{z8}). One easily obtains
\beq
J_1 = \bar g^4\frac{3}{\epsilon^2}\frac{\Gamma(1-2\epsilon)
\Gamma^3(1+\epsilon)}{\Gamma^2(1-\epsilon)\Gamma(1+3\epsilon)}~.
\label{z17}
\zen
The integrals $J_2$ and $J_3$ can be calculated in the same way  with the 
help of the representation 
\beq
% \ln a = -\frac{d}{d\nu}a^{-\nu}\arrowvert_{\nu=0}~.
\ln a = -\frac{d}{d\nu}a^{-\nu}|_{\nu=0}~.
\label{z18}
\zen
The result is
\beq
J_2 = J_1[\psi(1-\epsilon)+\psi(2\epsilon)-\psi(1-2\epsilon)-
\psi(3\epsilon)]~,
\label{z19}
\zen
\beq
J_3 = J_1[\psi(1)+\psi(\epsilon)-\psi(1-2\epsilon)-\psi(3\epsilon)]~.
\label{z20}
\zen
\vskip.03cm
The calculation of the integral $I_2$ is much more complicated and it 
appears that in the general case $D\neq 4$ it can be given only in terms of
infinite series.
\footnote{We are thankful to A.V. Kotikov who paid our 
attention on papers \cite{KK} where $I_2$ is expressed in terms of 
generalized hypergeometric series} 
Instead in the interesting for physical applications case $D\rightarrow 4$, 
the integral is expressed in terms of the 
well-known values of the Riemann zeta function. Because of the importance 
of its contribution we present the calculation of the integral in the 
following section. The result is
\beq
I_2  = \frac{\bar g^4}{\epsilon^2}
\left[-\frac{1}{\epsilon}+2\epsilon\psi'(1)-13\epsilon^2\psi''(1)\right]~.
\label{z21}
\zen
Let us remind that 
\beq
\psi'(1) = \zeta(2) = \frac{\pi^2}{6}~,~~~~~~\psi''(1) = -2\zeta(3) \approx
-2.404~.
\label{z22}
\zen
Expanding for the physical case $\epsilon \rightarrow 0$ the integrals 
$I_1$ and $J_1$-$J_3$, given respectively in Eqs. (\ref{z16}), (\ref{z17}), 
(\ref{z19}) and (\ref{z20}), and the coefficient of $I_1-2J_1$ in
Eq. (\ref{z11}) we find
\beq
I_1 = \frac{4\bar g^4}{\epsilon^2}[1-2\epsilon^2\psi'(1) -
2\epsilon^3\psi''(1)]~,
\label{z23}
\zen
\beq
J_1 = \frac{3\bar g^4}{\epsilon^2}\left[1-2\epsilon^2\psi'(1) -
5\epsilon^3\psi''(1)\right]~,
\label{z24}
\zen
\beq
J_2 = \frac{\bar g^4}{\epsilon^2}\left[-\frac{1}{2\epsilon}+
\epsilon\psi'(1) -\frac{19}{2}\epsilon^2\psi''(1)\right]~,
\label{z25}
\zen
\beq
J_3 = \frac{\bar g^4}{\epsilon^2}\left[-\frac{2}{\epsilon}+
4\epsilon\psi'(1) -8\epsilon^2\psi''(1)\right]~,
\label{z26}
\zen
and
\beqn
\psi(1)+2\psi(\epsilon)-\psi(1-\epsilon)-2\psi(1+2\epsilon)+\frac{1}
{(1+2\epsilon)}\left(\frac{1}{\epsilon}+\frac{1+\epsilon}{2(3+2\epsilon)}
\right) = 
\zenn
\beq
-\frac{1}{\epsilon}-\frac{11}{6}-\left(\psi'(1)-\frac{67}{18}\right)
\epsilon-\left(\frac{7}{2}\psi''(1)+\frac{202}{27}\right)\epsilon^2~.
\label{z27}
\zen
Finally, substituting the results (\ref{z21}), (\ref{z23})-(\ref{z27}) into 
Eq.(\ref{z11}), we obtain the gluon contribution to the two-loop 
correction to the gluon trajectory:
\beq
\omega^{(2)}_g(t) = \bar g^4(\vec q^{~2})^{2\mbox{\normalsize 
$\epsilon$}}\frac{1}{\epsilon^2}\left[\frac{11}{3}+\left(2\psi'(1)-
\frac{67}{9}\right)\epsilon+\left(\psi''(1)-\frac{22}{3}\psi'(1)+
\frac{404}{27}\right)\epsilon^2\right]~.
\label{z28}
\zen

\vskip 0.3cm

{\bf 3. CALCULATION OF THE INTEGRAL $I_2$}

We introduce the Feynman parameters $x_1$ and $x_2$ to join the
denominators depending on $q_1$ and $q_2$ correspondingly in Eq. (\ref{z12}),
and the parameter $z$ to join the results of these two parametrizations.
Then we integrate over $\vec q_2$ keeping the difference 
$\vec \Delta=\vec q_1-\vec q_2$ fixed and after applying the relation 
(\ref{z18}) arrive at
\beq
I_2 = \frac{\bar g^4}{\Gamma^2(1-\epsilon)}\int_0^1dz\int_0^1dx_1
\int_0^1dx_2\left(-\frac{\partial}{\partial \nu}\right)f(\nu,D)|_{\nu=0}~,
\label{z29}
\zen
where
\beq
f(\nu,D) = \frac{\Gamma\left(5-\frac{D}{2}\right)}{\pi^{\frac{D-2}{2}}}
(\vec q^{~2})^{(6-D+\nu)}z(1-z)
\label{z30}
\zen
\beqn
\times\int\frac{d^{D-2}\Delta}
{(\vec \Delta^2)^{\nu}[z(\vec \Delta^2+x_1\vec q^2-2x_1\vec \Delta\cdot 
\vec q)+(1-z)x_2\vec q^2-(z(\vec \Delta-x_1\vec q)-(1-z)x_2\vec q)^2]^
{5-\frac{D}{2}}}~.
\zenn
The integration over $\vec \Delta$ with the help of the generalized 
Feynman parametrization with parameter $x$ and the subsequent integration 
over $x$ by parts yield
\beq
f(\nu,D) = \frac{\Gamma(6-D+\nu)}{\Gamma(1+\nu)}(z(1-z))^{2-\frac{D}{2}+
\nu}\int_0^1dx(1-x)^{\nu}\frac{d}{dx}\frac{x^{\frac{D}{2}-2-\nu}}
{(Q(x))^{6-D+\nu}}
\label{z31}
\zen
with
\beq
Q(x) = zx_1(1-x_1)+(1-z)x_2(1-x_2)+z(1-z)(1-x)(x_1-x_2)^2~.
\label{z32}
\zen
It is convenient to present the derivative of $f(\nu,D)$ with respect to
$\nu$ in the form
\beq
-\frac{\partial}{\partial \nu}f(\nu,D)|_{\nu=0} = f_1+f_2+f_3~,
\label{z33}
\zen
where 
\beq
f_1 = \left(\psi(1)+\frac{\partial}{\partial D}\right)f(0,D)~,
\label{z34}
\zen
\beq
f_2 = -\frac{1}{2}\ln (z(1-z))f(0,D) = -\frac{\Gamma(6-D)}{2}
\frac{(z(1-z))^{2-\frac{D}{2}}}{(Q(1))^{6-D}}\ln (z(1-z))~,
\label{z35}
\zen
\beq
f_3 = -\Gamma(6-D)(z(1-z))^{2-\frac{D}{2}}\int_0^1dx\ln (1-x)
\frac{d}{dx}\frac{x^{\frac{D}{2}-2}}{(Q(x))^{6-D}}~.
\label{z36}
\zen
Let us denote the contributions of the functions $f_i$ to the integral 
$I_2$ in Eq. (\ref{z29}) as ${\cal I}_i$. It is easy to see that
\beqn
{\cal I}_1 = 
\zenn
\beq
\frac{\bar g^4}{\Gamma^2(1-\epsilon)}\left(\psi(1)+
\frac{1}{2}\frac{d}{d{\epsilon}}\right)
\frac{\Gamma^2(1-\epsilon)}{\bar g^4}I_1 = 
I_1[\psi(1)+2\psi(\epsilon)-\psi(1-\epsilon)-2\psi(2\epsilon)]~.
\label{z37}
\zen
\vskip.03cm
The calculation of ${\cal I}_2$ and ${\cal I}_3$ is not so simple. We will 
perform it in the limit $\epsilon \rightarrow 0$. The idea underlying the
calculation is to extract those regions of the integration range which lead to 
singular contributions for $\epsilon \rightarrow 0$ and to put $\epsilon=0$ in
the remaining region. Considering the integrand in Eq. (\ref{z12}) and our 
way of the Feynman parametrization it is easy to understand that the singular 
contributions can give only vicinities of points $x_i=0$ and $x_i=1$, with 
$i=1,2$. Taking into account the symmetry of the function $Q(x)$ of Eq. 
(\ref{z32}) under the simultaneous substitutions $x_1 \leftrightarrow 1-x_1$ 
and $x_2 \leftrightarrow 1-x_2$, we can restrict ourselves to the region 
$x_i>0$, $x_1+x_2<1$ taking its contribution twice. It is convenient to 
introduce new variables defined through
\beq
x_1 = \lambda t~,~~~~~~~~x_2 = (1-\lambda)t~.
\label{z38}
\zen
In these variables $Q(x)$ becomes
\beqn
Q(x) = tR(x)~,
\zenn
\beq
R(x) = z\lambda(1-\lambda t)+(1-z)(1-\lambda)(1-(1-\lambda)t)+(1-x)z(1-z)
(2\lambda-1)^2t~,
\label{z39}
\zen
and we have
\beq
{\cal I}_2 = -\frac{\bar g^4}{\Gamma^2(1-\epsilon)}\int_0^1dz(z(1-z))^
{-\epsilon}\ln(z(1-z))g(z)~,
\label{z40}
\zen
\beq
{\cal I}_3 =  -\frac{2\bar g^4}{\Gamma^2(1-\epsilon)}\int_0^1dz(z(1-z))^
{-\epsilon}\int_0^1dx\ln(1-x)g(x,z)~,
\label{z41}
\zen
where
\beq
g(z) = \Gamma(2-2\epsilon)\int_0^1dtt^{2\epsilon-1}\int_0^1d{\lambda}
\frac{1}{(R(1))^{2-2\epsilon}}~,
\label{z42}
\zen
\beq
g(x,z) = \Gamma(2-2\epsilon)\int_0^1dtt^{2\epsilon-1}\int_0^1d{\lambda}
\frac{d}{dx}\frac{x^{\epsilon}}{(R(x))^{2-2\epsilon}}~.
\label{z43}
\zen
The singular points are $t=0$, $\lambda=0$ and $\lambda=1$. Therefore, we 
divide the integration region in four pieces:
\beqn
a)~~~~0 < t < \delta~,~~~~0< \lambda < 1~;~~~~~~~~b)~~~~\delta < t < 1~,
~~~~0 < \lambda < \delta;
\zenn
\beq
c)~~~~\delta < t < 1~,~~~~0 < 1-\lambda < \delta~;~~~~~~~~d)~~~~
\delta < t < 1~,~~~~\delta < \lambda < 1-\delta~.
\label{z44}
\zen
Here we have introduced an intermediate parameter $\delta$ assuming that
\beq
e^{-\frac{1}{\epsilon}} \ll \delta \ll \epsilon^n
\label{z45}
\zen
for any fixed power n. Due to the first of these inequalities one can put 
$\epsilon=0$ in the region d) and due to the second one we can neglect $t$, 
$\lambda$ and $(1-\lambda)$ in comparison with 1 in the regions a), b) and 
c) correspondingly. Of course, $\delta$ is cancelled in the whole 
expression for $g(z)$ and $g(x,z)$.
\vskip.03cm
In the region a) we have independently on $x$
\beq
R = z\lambda+(1-z)(1-\lambda)
\label{z346}
\zen
so that the integration in Eqs. (\ref{z42}) and (\ref{z43}) is quite simple. 
Performing a suitable expansion in $\epsilon$ we find
\beq
g^{(a)}(z) = \Gamma(1-2\epsilon)\left[\frac{\delta^{2\epsilon}}
{2\epsilon}\left(z^{2\epsilon -1}+(1-z)^{2\epsilon -1}\right)+\frac{2}
{1-2z}\ln\left(\frac{z}{1-z}\right)\right]~,
\label{z47}
\zen
\beq
g^{(a)}(x,z) = \frac{\Gamma(1-2\epsilon)}{2}\delta^{2\epsilon}
x^{\epsilon-1}\left(z^{2\epsilon -1}+(1-z)^{2\epsilon -1}\right)~.
\label{z48}
\zen
Notice that in $g^{(a)}(z)$ we need to keep terms which were neglected 
in $g^{(a)}(x,z)$.
\vskip.03cm
In the region b) $R(x)$ takes the form
\beq
R(x) = \lambda+r(x)~,~~~~~~~~r(x) = (1-z)[1-t(1-z(1-x))]
\label{z49}
\zen
and considerably simplifies for $x=1$. It is convenient to start with 
the integration over $\lambda$ without deriving with respect to $x$ in 
$g(x,z)$. After that we can put $\epsilon=0$ in the contributions containing 
$\delta+r(x)$. The integration over $t$ in the case of $g(z)$ can be 
done without problems and gives
\beq
g^{(b)}(z) = \Gamma(1-2\epsilon)\left[(1-z)^{2
\mbox{\normalsize $\epsilon$}-1}\left(\frac
{\Gamma^2(2\epsilon)}{\Gamma(4\epsilon)}-\frac{\delta^{2
\mbox{\normalsize $\epsilon$}}}
{2\epsilon}\right)-\frac{1}{1-z+\delta}\ln\left(\frac{1-z+\delta}
{\delta^2}\right)\right]~.
\label{z50}
\zen
As for $g(x,z)$, using the approximate relation
\beq
t^{2\mbox{\normalsize $\epsilon$}-1}\frac{d}{dx}\frac{x^
{\mbox{\normalsize $\epsilon$}}}
{(r(x))^{1-2
\mbox{\normalsize $\epsilon$}}} = \frac{\epsilon}{x}\frac{t^{-1}}
{(1-z)^{1-2\mbox{\normalsize $\epsilon$}}(1-tx)}+
\frac{x^{\mbox{\normalsize $\epsilon$}}z(1+2\epsilon\ln t)}
{1-z(1-x)}\frac{d}{dt}(r(x))^{2\mbox{\normalsize $\epsilon$}-1}
\label{z51}
\zen
the integration over $t$ yields
\beqn
g^{(b)}(x,z) = \Gamma(1-2\epsilon)\left[\frac{(1-z)^{2
\mbox{\normalsize $\epsilon$}-1}zx^{\mbox{\normalsize $\epsilon$}}}
{(1-z(1-x))}\left((z(1-x))^{2\mbox{\normalsize $\epsilon$}-1}-1\right)\right.
\zenn
\beq
\left. + \frac{\epsilon}{x}(1-z)^{2\mbox{\normalsize $\epsilon$}-1}
\ln\left(\frac{1-x}{\delta}\right)+\frac{1}{1-z+\delta}\frac{\partial}
{\partial x}\ln(\delta + z(1-z)(1-x))\right]~.
\label{z52}
\zen
\vskip.03cm
The calculation of $g^{(c)}(z)$ and $g^{(c)}(x,z)$ is immediate. In fact, 
because of the symmetry of the function $R(x)$ in Eq. (\ref{z39}) under the 
replacement $z \leftrightarrow 1-z$ and $\lambda \leftrightarrow 1-\lambda$, 
these two contributions can be obtained from Eqs. (\ref{z50}) and 
(\ref{z52}) correspondingly by the substitution $z \leftrightarrow 1-z$.
\vskip.03cm
Finally we pass to the region d). After putting $\epsilon=0$ 
the integration over $t$ becomes quite straightforward. The subsequent 
integration over $\lambda$ is very simple in the case of $g(z)$ and gives
\beq
g^{(d)}(z) = \frac{1}{z+\delta}\ln\left(\frac{z+\delta}{\delta^2}\right)+
\frac{1}{1-z+\delta}\ln\left(\frac{1-z+\delta}{\delta^2}\right)
+\frac{2}{1-2z}\ln\left(\frac{1-z}{z}\right)~.
\label{z53}
\zen
For the case of $g(x,z)$ it is convenient to integrate without taking the 
derivative with respect to $x$. After integration over $t$ one can represent 
the coefficient in the term with the logarithm as
\beq
(z\lambda+(1-z)(1-\lambda))^{-2} = -\frac{d}{d\lambda}
\left(\frac{1-2\lambda}{z\lambda+(1-z)(1-\lambda)}\right)
\label{z54}
\zen 
and integrating by parts over $\lambda$ arrives at
\beq
g^{(d)}(x,z) = -\frac{\partial}{\partial x}\frac{\ln(\delta+z(1-z)(1-x))}
{(z+\delta)(1-z+\delta)}~.
\label{z55}
\zen
\vskip.03cm
Using the results given in Eqs. (\ref{z47}), (\ref{z50}) and (\ref{z53}) 
we obtain with the required accuracy
\beq
g(z) = g^{(a)}(z)+g^{(b)}(z)+g^{(b)}(1-z)+g^{(d)}(z) = 
\Gamma(1-2\epsilon)\frac{\Gamma^2(2\epsilon)}{\Gamma(4\epsilon)}
\left(z^{2\mbox{\normalsize $\epsilon$}-1}+(1-z)^
{2\mbox{\normalsize $\epsilon$}-1}\right)~.
\label{z56}
\zen
Analogously, from Eqs. (\ref{z48}), (\ref{z52}) and (\ref{z55}) for $g(x,z)$ 
we have
\beqn
g(x,z) = g^{(a)}(x,z)+g^{(b)}(x,z)+g^{(b)}(x,1-z)+g^{(d)}(x,z) = 
\zenn
\beqn
\left. \Gamma(1-2\epsilon)\left((1-z)^{2\mbox{\normalsize $\epsilon$}-1}z^
{2\mbox{\normalsize $\epsilon$}}+(1-z)^{2
\mbox{\normalsize $\epsilon$}}z^{2\mbox{\normalsize $\epsilon$}-1}\right)
\right.
\zenn
\beq
\times \left[x^{\mbox{\normalsize $\epsilon$}}(1-x)^{2
\mbox{\normalsize $\epsilon$}-1}+\frac{\epsilon}{x}
\left(3\ln(1-x)+\frac{1}{2}\ln x\right)+\frac{1}{2x}\right]~.
\label{z57}
\zen
\vskip.03cm
Substituting expressions (\ref{z56}) and (\ref{z57}) into Eqs. (\ref{z40}) 
and (\ref{z41}) respectively and performing the integration over $x$ and 
$z$ we obtain
\beqn
{\cal I}_2 = -2\bar g^4\frac{\Gamma(1-2\epsilon)}{\Gamma^2(1-\epsilon)}
\frac{\Gamma^2(2\epsilon)}{\Gamma(4\epsilon)}\frac{\Gamma(\epsilon)
\Gamma(1-\epsilon)}{\Gamma(1)}[\psi(\epsilon)+\psi(1-\epsilon)
-2\psi(1)] = 
\zenn
\beq
\frac{2\bar g^4}{\epsilon^2}\left[\frac{1}{\epsilon}-2\epsilon\psi'(1)-
10\epsilon^2\psi''(1)\right]~,
\label{z58}
\zen
and
\beqn
{\cal I}_3 = -4\bar g^4\frac{\Gamma(1-2\epsilon)}{\Gamma^2(1-\epsilon)}
\frac{\Gamma(\epsilon)\Gamma(1+\epsilon)}{\Gamma(1+2\epsilon)}
\zenn
\beqn
\times \left[\frac{\Gamma(2\epsilon)\Gamma(1+\epsilon)}{\Gamma(1+3\epsilon)}
(\psi(2\epsilon)-\psi(1+3\epsilon))-\frac{1}{2}\psi'(1)-\frac{13}{4}
\epsilon\psi''(1)\right] =
\zenn
\beq
\frac{\bar g^4}{\epsilon^2}\left[\frac{1}{\epsilon}+2\epsilon\psi'(1)+
13\epsilon^2\psi''(1)\right]~.
\label{z59}
\zen
Using last two equations and Eq. (\ref{z37}) with $I_1$ given by Eq. 
(\ref{z23}) we obtain for $I_2={\cal I}_1 +{\cal I}_2 +{\cal I}_3$ the result 
(\ref{z21}). 
\footnote{A.V. Kotikov informed us that he also obtained the same result 
from the generalized hypergeometric series \cite{KK} }

\vskip 0.3cm

{\bf 4. RENORMALIZATION AND DISCUSSION}
 
Up to now we worked with unrenormalized quantities. But renormalization is 
quite trivial. Since the trajectory itself must not be renormalized, we 
have only to use the renormalized coupling constant $g_{\mu}$ instead of 
the bare one $g$ . In the ${\overline {MS}}$ scheme one has
\beq
g=g_{\mu}\mu^{-\mbox{\normalsize $\epsilon$}}\left[1+\left(\frac{11}{3}-
\frac{2}{3}\frac{n_f}{N}\right)\frac{\bar g_{\mu}^2}{2\epsilon}\right]~,
\label{z60}
\zen
where
\beq
\bar g_{\mu}^2 = \frac{g_{\mu}^2N\Gamma(1-\epsilon)}{(4\pi)^{2+
\mbox{\normalsize $\epsilon$}}}~,
\label{z61}
\zen
and for the gluon trajectory in the two-loop approximation we obtain
\beqn
\omega(t) = -\bar g^2_{\mu}\left(\frac{\vec q^{~2}}{\mu ^2}\right)^
{\mbox{\normalsize $\epsilon$}}
\frac{2}{\epsilon}\left\{1+\frac{\bar g^2_{\mu}}{\epsilon}
\left[\left(\frac{11}{3}-\frac{2}{3}\frac{n_f}{N}\right)
\left(1-\frac{\pi^2}{6}\epsilon ^2\right)-\right.\right.
\zenn
\beqn
\left.\left. \left(\frac{\vec q^{~2}}{\mu ^2}\right)^
{\mbox{\normalsize $\epsilon$}}\left(\frac{11}{6}+\left(\frac{\pi^2}{6}-
\frac{67}{18}\right)
\epsilon+\left(\frac{202}{27}-\frac{11\pi^2}{18}-\zeta(3)\right)\epsilon^2-
\right.\right.\right.
\zenn
\beq
\left.\left.\left. \frac{n_f}{3N}\left(1-\frac{5}{3}\epsilon+\left(
\frac{28}{9}-\frac{\pi^2}{3}\right)\epsilon^2\right)\right)\right]\right\}~.
\label{z62}
\zen
The remarkable fact which occurred is the cancellation of the third order 
poles in $\epsilon$ existing in separate contributions to the gluon 
correction $\omega^{(2)}_g(t)$ (\ref{z11}). Possibly this fact indicates 
that the representation (\ref{z11}) is not the best one. From the other 
hand a simpler representation for $\omega^{(2)}_g(t)$ means that there 
should exist a simple expression for the integral $I_2$. In any case, the 
cancellation of the terms with $\epsilon^{-3}$ is very important for the 
absence of infrared divergences in the corrections to the BFKL equation. 
As the result of this cancellation the gluon and quark contributions to 
$\omega^{(2)}(t)$ have similar infrared behaviour. Moreover, the 
coefficient of the leading singularity in $\epsilon$ is proportional to 
the coefficient of the one-loop $\beta$ function. This means that infrared 
divergences are strongly correlated with ultraviolet ones. The 
correlation is unique in the sense that it provides the independence of 
singular contributions to $\omega(t)$ on $\vec q^2$. Indeed, expanding 
Eq. (\ref{z62}) we have
\beqn
\omega(t) = -\bar g^2_{\mu}\left(\frac{2}{\epsilon}+2\ln\left(
\frac{\vec q^{~2}}{\mu^2}\right)\right)-\bar g^4_{\mu}\left[\left(\frac{11}{3}
-\frac{2}{3}\frac{n_f}{N}\right)\left(\frac{1}{\epsilon^2}-\ln^2\left(
\frac{\vec q^2}{\mu^2}\right)\right)\right.
\zenn
\beq
\left. +\left(\frac{67}{9}-\frac{\pi^2}{3}-\frac{10}{9}\frac{n_f}{N}\right)
\left(\frac{1}{\epsilon}+2\ln\left(\frac{\vec q^2}{\mu^2}\right)\right)-
\frac{404}{27}+2\zeta(3)+\frac{56}{27}\frac{n_f}{N}\right]~.
\label{z63}
\zen
Eq. (\ref{z63}) exhibits explicitly all singularities of the trajectory in 
the two-loop approximation and gives its finite part in the limit 
$\epsilon \rightarrow 0$. Of course, the singularities should cancel
putting all the corrections into the BFKL equation. We hope to show this in 
a subsequent paper.

\vskip 1.5cm
\underline {Acknowledgement}: One of us (V.S.F.) thanks the Dipartimento di
Fisica della Universit\`a della Calabria and the Istituto Nazionale di 
Fisica Nucleare - Gruppo collegato di Cosenza for their warm hospitality 
while part of this work was done.

\end{document}